\documentclass[twocolumn,
preprintnumbers,prl,nofootinbib,
superscriptaddress
]{revtex4-1}

\usepackage{amsmath,amssymb}
\usepackage[usenames,dvipsnames,svgnames,table]{xcolor}
\usepackage[colorlinks,citecolor=DarkGreen,linkcolor=FireBrick,urlcolor=FireBrick,linktocpage]{hyperref}
\usepackage{graphicx}
\usepackage{times}

\def\Nequals#1{$\mathcal{N}{=}#1$}
\def\bR{\mathbb{R}}
\def\bZ{\mathbb{Z}}
\def\so{\mathfrak{so}}
\def\SO{\mathrm{SO}}

\def\Spin{\mathrm{Spin}}
\def\u{\mathfrak{u}}

\def\su{\mathfrak{su}}
\def\SU{\mathrm{SU}}
\def\e{\mathfrak{e}}

\def\tr{\mathop{\mathrm{tr}}}

\def\TMF{\mathrm{TMF}}
\def\sFL{\mathsf{F_L}}

\def\beq#1\eeq{\begin{align}#1\end{align}}
\def\d{\mathrm{d}}

\def\hete{(E_8 \times E_8) \rtimes \bZ_2}
\def\hets{\Spin(32)/\bZ_2}
\def\hetss{\dfrac{\Spin(32)}{\bZ_2}}

\begin{document}

\title{Non-supersymmetric heterotic branes}
%\preprint{IPMU-19-0111}

\author{Justin Kaidi}
\affiliation{Department of Physics, University of Washington, Seattle, WA, 98195, USA}
\author{Kantaro Ohmori}
\affiliation{Faculty of Science, University of Tokyo, Bunkyo, Tokyo 113-0033, Japan}
\author{Yuji Tachikawa}
\affiliation{Kavli Institute for the Physics and Mathematics of the Universe (WPI), \\
 University of Tokyo,  Kashiwa, Chiba 277-8583, Japan}
\author{Kazuya Yonekura}
\affiliation{Department of Physics, Tohoku University, Sendai, Miyagi 980-8578, Japan}

%\date{\today}

%-----------------------------------------
\begin{abstract}
The common statement that any consistent quantum gravity theory contains 
dynamical objects with all possible charges suggests that 
there are still a number of hitherto-unidentified branes in string theory.
Here we give the exact worldsheet 
description of  near-horizon limits of non-supersymmetric
 $p$-branes 
in ten-dimensional $\hets$ or $\hete$ heterotic superstring theories
for $p=7,6,4,0$. 
\end{abstract}

\maketitle

\section{Introduction}

It is often said that, in a consistent theory of quantum gravity,
there should be a dynamical object for every possible charge;
for a recent precise formulation, see \cite{McNamara:2019rup}.
A naive argument for this statement is based on the observation that nothing seems to forbid a semiclassical black hole or black brane solution from carrying a given charge.
The endpoint of Hawking evaporation starting from that semiclassical configuration should then provide the desired dynamical object. 
This, however, is often easier said than done. 

One successful example is the case of D-branes in ten-dimensional string theories.
The low-energy approximations to string theories were long known to contain abelian antisymmetric-tensor gauge fields with various numbers of spacetime indices,
known as Ramond-Ramond (RR) fields. 
When supersymmetric string theories were first found in the mid-1980s, 
no objects carrying these RR fields were known.
Then, in the early 1990s, black-brane solutions carrying RR charges were constructed in \cite{Horowitz:1991cd}, 
strongly suggesting the existence of more basic objects carrying these charges.
This eventually led to the recognition that D-branes, on which fundamental strings can end, carry RR charges \cite{Polchinski:1995mt}.

There remain a number of subtle charges in string theories, for which the corresponding branes are not known.
The aim of this letter is to provide a concrete construction of the worldsheet theories of four of them in the 
$G$ heterotic string theories, where $G$ is either $\hets$ or $\hete$.
We start with the case of a $\mathbb{Z}_2$-charged 6-brane in the $\hets$ theory, 
for which the aforementioned program of starting from a black brane solution and  
taking the extremal limit works best.
In fact, we will be able to give an exact worldsheet description of the extremal black 6-brane solution.

We will see that the analysis can be generalized to the cases of a 7-brane, a 4-brane, and a  
0-brane,
by replacing the angular part $S^2$ of the 6-brane solution 
with $S^1$, $S^4$, and $S^8$, respectively.
This provides us with  
exact worldsheet descriptions of near-horizon limits of $p$-branes in heterotic string theories where $p$ is of the form $8-2^i$, for $i=0,1,2,3$,
although we do not have a deep understanding behind this pattern.

These branes are based on gauge configurations on $S^{8-p}$ associated to some particular elements of the homotopy groups $\pi_{7-p}(G)$.
In particular, 
the $0$-brane describes an endpoint of the  
$\hets$ heterotic string, 
giving a concrete realization of the open heterotic strings discussed previously in \cite{Polchinski:2005bg}. Likewise, the $4$-brane has been discussed previously in \cite{Bergshoeff:2006bs} in the M-theory frame.
The $6$-brane and  the $7$-brane are new, to the authors' knowledge.
We also note that our branes in $\hets$ heterotic string theory, when dualized to Type I, are not D-branes.

Before proceeding, we note that a new non-supersymmetric 7-brane in F-theory was discussed in \cite{Dierigl:2022reg,Debray:2023yrs}. 
When combined with our results, it strongly suggests that there are many more non-supersymmetric branes waiting to be discovered in string/M-theory.

\section{The black 6-brane}
We start from the case of a $\bZ_2$ charge in the $\hets$ heterotic string theory.
This theory is known to have no dynamical particles in the vector representation of the Lie algebra $\so(32)$,
although there are particles in the adjoint and one of the spinor representations.
This means that the gauge group is a $\bZ_2$ quotient of $\Spin(32)$ which is not $\SO(32)$ \cite{Witten:1997bs}.
Correspondingly, there is a $\bZ_2$ magnetic charge associated to $\pi_1(\hets) \simeq \bZ_2$.
If there is a dynamical object carrying this $\bZ_2$ charge, 
it should be a 6-brane, whose angular part $S^2$ supports a magnetic flux of $\hets$ that is not compatible with the vector representation.

A concrete gauge configuration meeting this criterion can be given as follows.
Consider an $\so(32)$ magnetic flux
\beq
\frac{F_{\so(32)}}{2\pi} = \bigoplus_{i=1}^{16} \begin{pmatrix} 0 & q_i \\ -q_i & 0 \end{pmatrix} \cdot \frac{\epsilon_2}{4\pi} 
\eeq
where $\epsilon_2$ is the volume form on $S^2$ such that $\int_{S^2} \epsilon_2 =4\pi$, and $q_i \in \bR$ specifies the magnetic flux. 
The standard Dirac quantization argument shows that a particle in the vector representation is compatible
only if $q_i \in \bZ$. 
On the other hand, one can check that the adjoint and one of the spinor representations are compatible with $q_i \in \frac12 \bZ$
as long as $q_i-q_j \in \bZ$ and $\sum_{i=1}^{16} q_i \in 2\bZ$.
Any such configuration with $q_i\not\in \bZ$ has the nontrivial $\bZ_2$ charge.
In particular, we will consider $q_i = 1/2$ for all $i$, which preserves a $\u(16) \subset \so(32)$ symmetry.

We now write down a supergravity solution having this monopole flux. 
In fact, this was already done in \cite{Horowitz:1991cd} and we can simply quote the result.
Letting the magnetic flux $q_i$ be a free variable for the moment,
the black brane solution in the heterotic frame is given in Eq.~(22) of \cite{Horowitz:1991cd}.
After accounting for the precise coefficients in the supergravity action (e.g. \cite{Polchinski:1998rr}) it has the form
\beq
&\d s^2= - \frac{(1-\frac{r_+}{r})}{(1-\frac{r_-}{r})} \d t^2 + \d \vec x\, ^2 + \frac{\d r^2}{(1-\frac{r_+}{r})(1-\frac{r_-}{r})} + r^2 \d \Omega^2_2~,  \nonumber \\
& ~e^{-2\phi}  = g_s^{-2}(1-\frac{r_-}{r})~, \quad 
8r_+r_- = \alpha'\sum_{i=1}^{16} q_i^2   ~, 
\eeq
where the first line is the metric, $\phi$ is the dilaton, $g_s$ is the string coupling at infinity, and $\alpha'$ is the standard string theory parameter.
The mass density of the brane is proportional to $ r_+ $ \cite{Horowitz:1991cd}.
%and is bounded below by $(\sum_{i=1}^{16} q_i^2)^{1/2}$.
This is a perfectly smooth, weakly-coupled semiclassical solution outside of the outer horizon when $r_+$ is large.

We now take the extremal near-horizon limit
by setting $r_0:=r_+ \to  r_-$ and $e^{y/r_0} := (r-r_0)/r_0 \to 0$, keeping $q_i$ fixed. This gives the following metric and the dilaton profile,
\beq
 \d s^2=\d x^\mu \d x_\mu + \d y^2 + r^2_0 \d \Omega^2_2~,  \quad
e^{-2\phi}  = g_s^{-2} e^{y/r_0}~. \label{extremal6}
\eeq
We see that the throat is infinitely long, and the $S^2$ part has constant size.
If we set $q_i=1/2$ for all $i$ as discussed above, 
we see that the size $r_0=(\alpha'/2)^{1/2}$ is stringy
and necessitates a worldsheet analysis, which we will provide momentarily.

\section{The general pattern}
Before describing the worldsheet theory of the 6-brane, let us outline the general pattern
which will appear for all of the branes studied in this letter. Let $G$ be either $\hets$ or $\hete$,
and let $G_1$ be the left-moving current algebra theory at level 1.
As is basically known (see e.g.~\cite{Classification} for a recent discussion and historical information), 
there are equivalences of theories (possibly up to discrete theta terms)
\beq
G_1 = [   H_k \times \SO(n)_1 ]/(-1)^{\sFL}~,   \label{eq:decomposition}
\eeq
where $ H$ is a group, 
$H_k$ is its current algebra theory at level $k$,
and $\SO(n)_1$ denotes $n$ Majorana-Weyl fermions $\psi_L^i~(i=1,\cdots, n)$. 
In particular, we define $\SO(1)_1$ to be a single Majorana-Weyl fermion.
We list the $G, H, n$, and $k$ that we use  in Table~\ref{table:decomposition}.
When $k =1$, the weight lattice of $H$, used as a Narain lattice, is self-dual but not even.
Both $H_k$ and $\SO(n)_1$ are spin-CFTs, 
and the notation $/(-1)^{\sFL}$ in Eq.~\eqref{eq:decomposition} denotes gauging by fermion parity $(-1)^{\sFL}$.
One can check that the central charges on both sides agree.
We remark that states of the theory $H_k$ (resp.~$\SO(n)_1$) in the NS sector transform in genuine representations of 
$H$ (resp.~$\SO(n)$), while states in the R sector may transform in projective representations
due to mixed anomalies between $H$ (resp.~$\SO(n)$) and $(-1)^{\sFL}$.

\begin{table}[t]
\caption{The groups and parameters in the equivalence Eq.~\eqref{eq:decomposition} which are used in this letter. The $p$-brane solution involves $n=8-p$.}
\label{table:decomposition}
\centering
\begin{tabular}{c||c|c|c|c}
$G$ \vphantom{$\bigm(\over \bigm($} &$\hete$&$\hetss$&$\hete$&$\hetss$   \\ 
\hline
\hline
$H$ \vphantom{$\bigm(\over \bigm($} &$E_8$&$\dfrac{\SU(16)}{\bZ_4}$&$\dfrac{E_7 \times E_7}{\bZ_2}$&$\dfrac{\Spin(24)}{\bZ_2}$ \\
\hline
$(n,k)$ &$(1,2)$&$(2,1)$&$(4,1)$&$(8,1)$  
\end{tabular}
\end{table}

For a $p$-brane with $p=8-n$, the angular directions form $S^n$.
Our gauge configurations for $G$ on $S^{n}$, discussed in detail later, will be such that the vector bundle of $\SO(n)$
is equal to the tangent bundle of $S^n$. (For $n=1$ the situation is more subtle, but the claim still holds.) 
The $S^n$ factor gives \Nequals{(0,1)} supermultiplets $(\phi^i,\psi_R^i)$, where $\phi^i$ are the bosonic coordinates of $S^n$
and $\psi_R^i$ are the right-moving superpartners,
and by coupling $\SO(n)_1=\{\psi_L^i \}$ to the tangent bundle we get \Nequals{(1,1)}
$S^n$ sigma model multiplets $(\phi^i,\psi_R^i,\psi_L^i)$, if we neglect the radial direction.

The semiclassical analysis of \Nequals{(1,1)} sigma models with target space $X$ in the RR sector gives a supersymmetric vacuum
for each basis element of the de~Rham cohomology groups $\bigoplus_{i=0}^{\dim X} H^i_\textrm{dR}(X)$,
represented as harmonic forms~ \cite{Witten:1982df}.
Moreover, $(-1)^{\sFL}$ acts as the Hodge star. For $X=S^n$, there are two vacua corresponding to 
$H_\textrm{dR}^0(S^n) \simeq  H_\textrm{dR}^n(S^n) \simeq {\mathbb R}$,
and they are exchanged by $(-1)^{\sFL}$. Thus the gauging by $(-1)^{\sFL}$ in Eq.~\eqref{eq:decomposition} has the effect
of identifying the two vacua, leaving a single vacuum. 
We assume that the vacuum of the \Nequals{(1,1)} $S^n$ sigma model is gapped in the IR, which is believed to be the case
for $n>1$, but will require additional relevant deformations for $n=1$.
The radial direction with an \Nequals{(0,1)} multiplet is coupled to the angular direction, but we assume that this coupling does not spoil the qualitative behavior 
of the \Nequals{(1,1)} $S^n$ part.
The simplest possibility consistent with worldsheet 't~Hooft anomaly matching (including
the central charge) after gapping the \Nequals{(1,1)} $S^n$ multiplets is then
\beq
\mathbb{R}^{p,1}  \times \mathbb{R}_\textrm{linear dilaton} \times H_k \label{eq:exactw}
\eeq
where $\mathbb{R}^{p,1} $ is the free \Nequals{(0,1)} sigma model parallel to the brane (which is the $x^\mu$ of Eq.~\eqref{extremal6} 
in the case of the $6$-brane), 
$ \mathbb{R}_\textrm{linear dilaton} $ is an \Nequals{(0,1)} linear dilaton in the radial direction (which is the $y$ of Eq.~\eqref{extremal6}), 
and $H_k$ is the spin-CFT appearing in Eq.~\eqref{eq:decomposition}. This is the worldsheet theory for the throat region of the $p$-brane.

\section{The 6-brane}
Let us now come back to the 6-brane. 
The left-moving current algebra of the $\hets$ theory has a realization in terms of sixteen complex Weyl fermions 
$\Psi_L^{1,\ldots,16}$ gauged by left-moving fermion parity $(-1)^{\sFL}$.
Our choice of monopole background on $S^2$ gives flux $q_i=1/2$ for each of the Weyl fermions, made consistent by the $(-1)^{\sFL}$ gauging.
We now use the worldsheet duality \begin{equation}
\Psi_L^{1,\ldots,16} / (-1)^{\sFL}
= [(\SU(16)/\bZ_4)_1 \times \Psi_L] /(-1)^{\sFL}~,
\label{dual-6}
\end{equation} where $(\SU(16)/\bZ_4)_1 $ (which is the part $H_k$ in Eq.~\eqref{eq:decomposition}) stands for the modular-invariant spin-CFT with $c_L=15$ based on this current algebra, 
and $\Psi_L$ (which is the part $\SO(n)_1$ with $n=2$) is a complex Weyl fermion.
In this alternative description, the magnetic flux for $\Psi_L$ is $2$, chosen to match the level of the associated $\u(1)$ current algebras: $16\cdot (1/2)^2 = 2^2$. A monopole background of charge $2$ is equivalent to the tangent bundle of $S^2$, regarded as a complex manifold $\mathbb{CP}^1$.
We thus get an  \Nequals{(1,1)} $S^2$ sigma model as discussed above, which is gapped. 
In the IR, we obtain the theory in Eq.~\eqref{eq:exactw}
with $p=6$ and $H_k = (\SU(16)/\bZ_4)_1$.

\section{The 4-brane}

The exact worldsheet description of the extremal solution can be generalized to a few other cases. 
We start with a 4-brane in the $\hete$ heterotic string theory. 
The associated charge is measurable on the angular $S^4$ surrounding the 4-brane in the following manner. 
Recall the relation $\d H \propto (\tr F^2)_1 + (\tr F^2)_2$, 
where the right-hand side represents the instanton numbers of the two $E_8$ factors and $H$ is the 3-form field strength of the $B$-field.
This relation prevents us from putting a single instanton on $S^4$.
However, we are still allowed to put an instanton of charge $1$ in one $E_8$ factor
and an instanton of charge $-1$ in the other $E_8$ factor, such that the $B$-field magnetic charges cancel.
This means that there is still a type of instanton charge measurable on $S^4$.

Such a configuration was previously studied in \cite{Bergshoeff:2006bs},
where non-extremal supergravity solutions were studied. 
Here we show that the extremal solution and its infrared limit can be found analogously to the case of the 6-brane.
Recall that there is a subalgebra $\su(2) \times \e_7 \subset \e_8$.
The crucial relation is the equality of the worldsheet theories \begin{equation}
(\hete)_1 = [  (\frac{E_7 \times E_7}{\bZ_2})_1  \times \psi_L^{1,2,3,4} ] /(-1)^{\sFL}~,\label{dual-4}
\end{equation}
where $((E_7 \times E_7)/\bZ_2)_1  $ is the modular-invariant spin-CFT based on this current algebra, 
and $\psi_L^{1,2,3,4} $ are four Majorana-Weyl fermions, i.e. $ \SO(4)_1$. 
We choose to put instantons in $\su(2) \times \su(2) \subset \e_8 \times \e_8$ with charge $(+1,-1)$.
The bundle in the vector representation of $\so(4) \simeq \su(2) \times \su(2)$ is 
the same as the tangent bundle of $S^4$. 
%For instance, the Euler number of $S^4$ is identified as $(+1)-(-1)=2$ in terms of the instanton charge $(+1,-1)$.
Thus we get an  \Nequals{(1,1)} $S^4$ sigma model, which is gapped, and consequently in the IR we obtain the theory in Eq.~\eqref{eq:exactw}
where $p=4$ and $H_k = ((E_7 \times E_7)/\bZ_2)_1 $.

\section{The 0-brane}

We next discuss a 0-brane. %Here we consider only one out of a few possibilities, leaving others to a future work \cite{FollowUp}. 
We focus on the $\hets$ theory.
On the angular $S^8$ directions  surrounding a 0-brane,
we can measure a charge $ \propto \int_{S^8}  \tr F^4$ associated to $\pi_7(\Spin(32)/\bZ_2)$.
To have a worldsheet description of such a brane, we use the equality \begin{equation}
(\hetss)_1 = [(\frac{\Spin(24)}{\bZ_2})_1 \times \psi_L^{1,\ldots,8}]/(-1)^{\sFL}~, \label{dual-0}
\end{equation} 
More precisely, this duality is realized as follows. First, there is a duality
$  (\psi_L^{1,\ldots,8}) = (\tilde\psi_L^{1,\ldots,8})/(-1)^{\mathsf{F_{L,1}}} $ of eight Majorana-Weyl fermions.
We also have 
$(\Spin(24)/\bZ_2)_1 = (\tilde\psi_L^{9,\ldots,32})/(-1)^{\mathsf{F_{L,2}}}  $. 
The gauging by $(-1)^{\sFL}$ in Eq.~\eqref{dual-0} 
has the effect of setting
$(-1)^{\mathsf{F_{L,1}}} = (-1)^{\mathsf{F_{L,2}}}$, and hence the right-hand side is $(\tilde\psi_L^{1,\ldots,32})/(-1)^{\sFL} $ which is $ (\Spin(32)/\bZ_2)_1$.

In the relation $  (\psi_L^{1,\ldots,8}) = (\tilde\psi_L^{1,\ldots,8})/(-1)^{\mathsf{F_{L,1}}} $,
the vector representation of one side is dual to one of the spinor representations of the other.
Now, if we take $\tilde\psi_L^{1,\ldots,8}$ to be in the spin bundle of $S^8$ which has a nonzero, unit value for
the charge measured by $ \int_{S^8} \tr F^4$,
then $\psi_L^{1,\ldots,8}$ is in the tangent bundle of $S^8$. 
Thus we get the \Nequals{(1,1)} $S^8$ sigma model which is gapped, and we obtain the theory in Eq.~\eqref{eq:exactw}
where $p=0$ and $H_k =(\Spin(24)/\bZ_2)_1$.

This system has a spacetime gravitational anomaly due to spacetime chiral fermions transforming in the fundamental of $\so(24)$.
Cancelling it requires the introduction of a fundamental string going into the throat of the 0-brane,
meaning that a fundamental string ends on this 0-brane, as discussed in \cite{Polchinski:2005bg}.
This can also be understood from the Green-Schwarz coupling $B \wedge \tr F^4$. The $\tr F^4$ flux effectively creates a $B$-field electric charge, which should be cancelled by a fundamental string insertion.

\section{The 7-brane}

Our final example is a 7-brane in the $\hete$ theory.
The charge measurable on the angular $S^1$ around the 7-brane is the possible exchange of the two $E_8$ factors, 
which is the meaning of the $\bZ_2$ 
part of the group $\hete$.
To analyze this system, we start from the equality (up to a discrete theta term) \begin{equation}
(\hete)_1 = [ (E_8)_2 \times \psi_L ] /(-1)^{\sFL}~, \label{dual-7}
\end{equation}
where $(E_8)_2$ is a purely left-moving spin-CFT based on this current algebra,
and $\psi_L$ is a single Majorana-Weyl fermion. 

There are some differences between this and the previous cases. One is that the \Nequals{(1,1)} $S^1$ sigma model is itself conformal, not gapped.
Another difference is that the nontrivial fundamental group of $S^1$ means that
there are different versions of  the theory %\eqref{worldsheet-uv-7} 
distinguished by the spacetime spin structure around $S^1$,
and by the existence or absence of a $\bZ_2$ holonomy around $S^1$.

We choose the NS spin structure and nontrivial $\bZ_2$ holonomy.
As we will detail in \cite{WithVivek}, 
these two choices conspire to make the combination $\psi_L e^{iw\tilde X}$
invariant under $(-1)^{\sFL}$ when $w$ is odd, where $e^{iw\tilde X}$ is the operator with winding number $w$.
When the radius of $S^1$ is made very small, these operators %(or more precisely their superpartners which can appear in the Lagrangian) 
become relevant,
corresponding to winding tachyons in the spacetime.
Activating the tachyon, for instance with $w=\pm 1$, means including the \Nequals{(0,1)} superpotential term 
$\int \d\theta_R\, \psi_L \cos \tilde X$,
and in particular the potential $V \sim ( \cos \tilde X )^2$ for the T-dual variable $\tilde X$.
This gives two gapped vacua at $\tilde X = \pm \pi/2$ exchanged by $(-1)^{\sFL} : \tilde X \mapsto \tilde X +\pi$.
(A more general analysis requires the mod 2 index.)
Thus we again obtain the theory in Eq.~\eqref{eq:exactw},
where now $p=7$ and $H_k =(E_8)_2$.

\section{Summary and remarks}

Let us summarize the branes  described so far:
\begin{itemize}
\item The 7-brane is in the $\hete$ heterotic string and carries a $\bZ_2$ charge capturing the potential exchange of the two $E_8$ factors on the angular $S^1$ direction. It is associated to $\pi_0(\hete) \simeq \bZ_2$.
Choosing the spacetime NS spin structure around $S^1$,
we can activate a tachyon.

\item The 6-brane is in the $\hets$ heterotic string, 
embeds the $\u(1)$ curvature of the tangent bundle of $S^2$ into $\u(1) \subset \u(16) \subset \so(32)$,
and carries a $\bZ_2$ charge associated to $\pi_1(\hets) \simeq \bZ_2$. 

\item The 4-brane is in the $\hete$ heterotic string,
embeds the $\so(4)\simeq \su(2)\times\su(2)$ curvature of the tangent bundle of $S^4$ into $\e_8\times \e_8$,
and carries the instanton charge $(+1,-1) \in \bZ \times \bZ \simeq \pi_3(\hete)$ on the angular $S^4$ directions.

\item The 0-brane is in the $\hets$ heterotic string,
embeds the $\so(8)$ curvature of the spin bundle of $S^8$ into $\so(32)$ %using a duality between the vector and one of the spinor representations,
and carries unit charge $1 \in \bZ \simeq \pi_7(\hets)$ measured by $\int_{S^8} \tr F^4$ on the angular $S^8$ directions.
Spacetime gravitational anomaly cancellation requires a fundamental string to end on it.

\end{itemize}

The worldsheet theories describing the throat region of the branes are given by Eq.~\eqref{eq:exactw} and Table~\ref{table:decomposition}.
In all cases, the angular part $S^{n}$ surrounding the $p$-brane (where $n=8-p$) is combined with some part of the internal current algebra theory $G_1$ 
and gapped, leaving a smaller current algebra theory $H_k$ based on the group $H$ at level $k$ listed 
in Table~\ref{table:decomposition},  as well as $\bR^{p,1}$ and a super linear dilaton $\mathbb{R}_\textrm{linear dilaton}$.

We end this letter with a couple of remarks:
\begin{itemize}
\item The same exact worldsheet descriptions, of the form $\bR^{1,p}$ times a super linear dilaton times a current algebra theory $H_k$, 
were previously found in the context of stable vacua of 10d tachyonic heterotic strings in \cite{Seiberg:2005nk,Hellerman:2007zz,Kaidi:2020jla}.
In particular, in \cite{Seiberg:2005nk} where the case $p=0$ was studied,
some distant relation to the open heterotic string of \cite{Polchinski:2005bg} was mentioned.
Here we find that it actually provides an exact infrared worldsheet description of the 0-brane on which an open heterotic string  can end.

\item As these systems do not preserve supersymmetry, 
and as the string coupling diverges at the core,
it is possible that the actual objects look qualitatively different from the ones described by the above worldsheet theories
(e.g. the possibility of spontaneous breaking of the $H$ symmetry at the core).
The existence of some branes is very likely if they can be obtained from black branes of finite mass density, as in the case of the 6-brane,
and there is topologically no way via bordism \cite{McNamara:2019rup} for the brane to dissipate into other fields (e.g. NS5 branes can dissipate into gauge instantons of large size,
although in this example they are energetically the same by BPS saturation). 
Finding descriptions of these branes in dual frames would be ideal.
We note that our 4-brane was realized in terms of M5-branes in \cite{Bergshoeff:2006bs}.
A discussion of the other branes in various duality frames will be presented in \cite{FollowUp}.

\item 

The branes discussed here also play important roles in the study of general heterotic string theories using the theory of topological modular forms as in \cite{Tachikawa:2021mby}.
In particular, the $p$-branes we have discussed above define special elements of the group $\TMF^{24+p}(\textrm{point})$,
and the subtle charges associated to them can also be understood in this context,
as will be discussed in \cite{WithMayuko}.

\end{itemize}

The authors plan to provide more details of these branes in the not-so-distant future,
from various points of view \cite{FollowUp}.

\section{Acknowledgments} 
The authors thank Yasunori Lee and Vivek Saxena for the collaboration at an early stage of the work, 
and Mayuko Yamashita for discussions on mathematical aspects. 

KO is supported by JSPS KAKENHI Grant-in-Aid No.22K13969 and the Simons Collaboration on Global Categorical Symmetries.
YT is supported in part  
by WPI Initiative, MEXT, Japan at Kavli IPMU, the University of Tokyo
and by JSPS KAKENHI Grant-in-Aid (Kiban-S), No.16H06335.
KY is supported in part by JST FOREST Program (Grant Number JPMJFR2030, Japan), MEXT-JSPS Grant-in-Aid for Transformative Research Areas (A) ``Extreme Universe'' (No. 21H05188), and JSPS KAKENHI (17K14265). 

\bibliographystyle{ytphys}
\bibliography{ref}

\end{document}